\documentclass[aps,prb, reprint, amsmath, amssymb, groupedaddress, superscriptaddress, showkeys,floatfix]{revtex4-2}

\usepackage{amsmath}
\usepackage{amssymb}
\usepackage{multirow}
\usepackage{graphicx}
\usepackage{siunitx}
\usepackage{dcolumn}
\usepackage{bm}
\usepackage{txfonts}
\usepackage[dvipsnames]{xcolor}
\usepackage[normalem]{ulem}
\usepackage{siunitx}
\usepackage{soul}
\setulcolor{red}
\usepackage[linkcolor=blue,urlcolor=blue,citecolor=blue,colorlinks=true]{hyperref}

\begin{document}

\title{
Elucidating Na$_2$KSb band structure: near-band-gap photoemission spectroscopy and DFT calculations
}

\author{S. A. Rozhkov}
\email{rozhkovs@isp.nsc.ru}
\affiliation{Rzhanov Institute of Semiconductor Physics, Siberian Branch,
Russian Academy of Sciences, Novosibirsk 630090, Russia}
\affiliation{Novosibirsk State University, Novosibirsk 630090, Russia}

\author{V. V. Bakin}
\affiliation{Rzhanov Institute of Semiconductor Physics, Siberian Branch,
Russian Academy of Sciences, Novosibirsk 630090, Russia}

\author{S. V. Eremeev}
\affiliation{Institute of Strength Physics and Materials Science, Russian Academy of Sciences, 634055 Tomsk, Russia}

\author{V. S. Rusetsky}
\affiliation{Rzhanov Institute of Semiconductor Physics, Siberian Branch,
Russian Academy of Sciences, Novosibirsk 630090, Russia}
\affiliation{JSC ``EKRAN FEP", Novosibirsk 630060, Russia}

\author{V. A. Golyashov}
\affiliation{Rzhanov Institute of Semiconductor Physics, Siberian Branch,
Russian Academy of Sciences, Novosibirsk 630090, Russia}
\affiliation{Novosibirsk State University, Novosibirsk 630090, Russia}
\affiliation{Synchrotron Radiation Facility SKIF, Boreskov Institute of
Catalysis, Siberian Branch, Russian Academy of Sciences, Koltsovo 630559, Russia}

\author{D. A. Kustov}
\affiliation{Rzhanov Institute of Semiconductor Physics, Siberian Branch,
Russian Academy of Sciences, Novosibirsk 630090, Russia}

\author{D. K. Orekhov}
\affiliation{Rzhanov Institute of Semiconductor Physics, Siberian Branch,
Russian Academy of Sciences, Novosibirsk 630090, Russia}
\affiliation{Novosibirsk State Technical University, Novosibirsk 630073, Russia}

\author{H. E. Scheibler}
\affiliation{Rzhanov Institute of Semiconductor Physics, Siberian Branch,
Russian Academy of Sciences, Novosibirsk 630090, Russia}
\affiliation{Novosibirsk State University, Novosibirsk 630090, Russia}
\affiliation{Novosibirsk State Technical University, Novosibirsk 630073, Russia}

\author{V. L. Alperovich}
\affiliation{Rzhanov Institute of Semiconductor Physics, Siberian Branch,
Russian Academy of Sciences, Novosibirsk 630090, Russia}
\affiliation{Novosibirsk State University, Novosibirsk 630090, Russia}

\author{O. E. Tereshchenko}
\email{teresh@isp.nsc.ru}
\affiliation{Rzhanov Institute of Semiconductor Physics, Siberian Branch,
Russian Academy of Sciences, Novosibirsk 630090, Russia}
\affiliation{Novosibirsk State University, Novosibirsk 630090, Russia}
\affiliation{Synchrotron Radiation Facility SKIF, Boreskov Institute of
Catalysis, Siberian Branch, Russian Academy of Sciences, Koltsovo 630559, Russia}

\date{\today}

\begin{abstract}

The electronic band structure of Na$_{2}$KSb was studied by a combination of low-energy photoemission spectroscopy and density functional theory (DFT) calculations. The optical and photoemission quantum efficiency (QE) spectra, along with longitudinal energy distribution curves (EDCs) of multialkali Na$_{2}$KSb(Cs,Sb) photocathodes were measured in the temperature range of 80--295\,K. The thresholds of various band-to-band transition in Na$_{2}$KSb were observed in the optical and QE spectra of Na$_{2}$KSb(Cs,Sb) photocathodes. The evolution of EDC derivatives with varying photon energy reveals a fine structure related to the emission of two types of electrons: (i) ballistic electrons, which are excited from heavy hole, light hole and split-off valence bands, and (ii) photoelectrons, that are captured in the side valleys of Na$_{2}$KSb conduction band. The analysis of EDCs and QE spectra allowed us to determine the band structure parameters of Na$_{2}$KSb at $T = 80$\,K, including the band gap $E_{\text{g}} = 1.52 \pm 0.02$\,eV, spin-orbit splitting $\Delta_{\text{SO}} = 0.59 \pm 0.04$\,eV and the energy separations between $\Gamma$ and side valleys of the conduction band: $\Delta_{\Gamma-\text{X}1} = 0.41 \pm 0.05$\,eV and $\Delta_{\Gamma-\text{X}2} = 0.65 \pm 0.05$\,eV. The experimentally determined band gaps and side valley positions, as well as the energies of the final electronic states of optical transitions are in good agreement with the DFT calculations. The obtained data on the hot electron dynamics and electronic band structure of Na$_{2}$KSb are crucial to improve the understanding of the photoemission processes in this material and will contribute to the development of the robust spin-polarized electron sources with multialkali photocathodes.

\end{abstract}

\maketitle

\section{\label{Sec.1}Introduction}

In recent years, the electronic \cite{Murtaza2016, Yalameha2018, Cocchi2019, Sharma2019, Cocchi2019-2, Khan2021, Sabnick2021, Amador2021, Wang2022, Wu2023, Schier2025} and vibrational \cite{Zhong2021, Singh2022, Yue2022} properties of alkali antimonide semiconductor compounds ($X_2Y$Sb; where $X$ and $Y$ = Li, Na, K, Cs, Rb) have attracted considerable attention due to their use in the accelerator facilities as electron sources \cite{Dunham2013, Musumeci2018, Petrushina2020, Guo2025}, as well as in thermoelectric \cite{Sharma2019, Singh2022, Yue2022} and photovoltaic \cite{Rusetsky2021, Wu2023} devices. Electron sources require robust photocathodes with high quantum efficiency (QE) and electron spin polarization, fast response time, as well as low mean transverse energy (MTE), which provides a high brightness. The use of alkali antimonide compounds as photocathodes provides an optimal combination of these key parameters, compared, on the one hand, to robust and fast, but very low-efficient and non-polarized metal photocathodes \cite{Dowell2009}, and, on the other hand, to III-V semiconductor photocathodes with negative effective electron affinity (NEA) \cite{Bell1973}, which have a record-high QE (up to 0.7) \cite{Uchiyama2005} and the ability to generate spin-polarized electrons \cite{Pierce1975}, but a slower response and require extreme-high vacuum for their operation \cite{Chanlek2014, Dunham2013}.

\begin{figure*}
\includegraphics[width=1.0\linewidth]{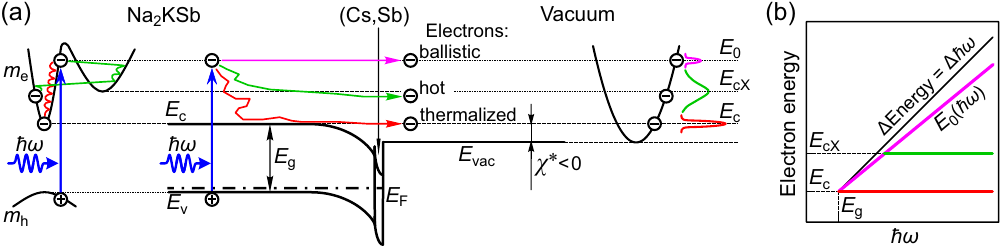}
\caption{\label{Fig.1} 
(a) Schematic energy band diagram of the Na$_{2}$KSb(Cs,Sb) photocathode with NEA and the photoemission pathways for ballistic, hot and thermalized electrons. $E_{\text{c}}$ and $E_{\text{v}}$ are the conduction band bottom and the valence band top in Na$_{2}$KSb, respectively, $E_{\text{g}}$ is the band gap, $E_{\text{vac}}$ is the vacuum level, $\chi^{*} = E_{\text{c}} - E_{\text{vac}}$ is the effective electron affinity, $E_{\text{F}}$ is the Fermi level, $E_{\text{0}}$ is the energy of ballistic electrons, $E_{\text{c}\text{X}}$ is the bottom of the conduction band side valley, $m_{\text{e}}$ and $m_{\text{h}}$ are the effective masses of electrons and holes, respectively. (b) The illustration of the dependence of characteristic electron energies on the excitation photon energy $\hbar\omega$.
}
\end{figure*}

Among various alkali antimonides, the bialkali cubic Na$_{2}$KSb compound with a direct band gap is the most suitable for high-QE (with QE up to 0.1) photocathodes, which are widely used in near-infrared low-light detection devices and robust high-efficiency electron sources at accelerator facilities \cite{Cultrera2016, Zhou2023, Dube2025}. The advantage of Na$_{2}$KSb is due to the opportunity to activate its surface to the state of NEA by the co-adsorption of Cs and Sb \cite{Spicer1958, Dolizy1982, Erjavec1997}. A recent discovery of high spin polarization \cite{Rusetsky2022} and low MTE values of electrons emitted from Na$_{2}$KSb(Cs,Sb) photocathodes \cite{Rozhkov2024,Rozhkov2025} proves the high potential of these photocathodes as robust high-brightness spin-polarized electron sources. As in the case of the III-V semiconductor NEA photocathodes \cite{Drouhin1985_1, Bell1973}, further development and optimization of Na$_{2}$KSb(Cs,Sb) photocathodes require the detailed knowledge on the electronic band structure of an active Na$_{2}$KSb layer.

The electronic band structure of Na$_{2}$KSb was calculated earlier by \textit{ab initio} methods in a number of papers \cite{Ettema2000, Murtaza2016, Yalameha2018, Sharma2019, Khan2021, Amador2021, Rusetsky2022}; however, the findings are inconsistent, and only part of them accounted for the spin-orbit interaction \cite{Yalameha2018, Sharma2019, Amador2021, Rusetsky2022}. The experimental data on the band structure of Na$_{2}$KSb are scarce and also contradictory. So far, the band gap $E_{\text{g}}$ is the only experimentally determined parameter of the Na$_{2}$KSb band structure. Moreover, the values of $E_{\text{g}}$ at room temperature $T = 295$\,K reported in literature varied from 1.0\,eV \cite{Spicer1958, Ghosh1978} to 1.4\,eV \cite{Hoene1972, Beguchev1993}. Recently, the combined experimental studies of optical and photoemission properties of high-efficiency Na$_{2}$KSb(Cs,Sb) photocathodes have confirmed the fundamental band gap value $E_{\text{g}} \approx 1.4$\,eV at $T = 295$\,K \cite{Rusetsky2021, Rozhkov2024}, as well as the electronic bands ordering and wave functions symmetry at the center of the Na$_{2}$KSb Brillouin zone \cite{Rusetsky2022}. Other parameters, such as valence band spin-orbit splitting, effective masses of electrons and holes, side valley positions of the conduction band (CB), which are crucial for the dynamics of non-equilibrium spin-polarized electrons upon photoemission, were not determined experimentally. The main reason for the scarcity of experimental data is the chemical instability of alkali antimonides in the air \cite{Schmeiber2018}. Thus, their growth (on foreign substrates) and subsequent studies should be carried out \textit{in situ}, in the same vacuum set-up, ensuring that the samples are not exposed to the air. This hinders the determination of the chemical composition, atomic structure and electronic properties of alkali antimonide films.

The state of NEA on the surface of Na$_{2}$KSb(Cs,Sb) photocathodes allows one to explore their near-band-gap photoemission spectroscopy to probe the electronic states in the conduction band of Na$_{2}$KSb, as well as the electron dynamics during photoemission \cite{James1969, Drouhin1985_1, Peretti1991, Orlov2000, Piccardo2014, Rozhkov2016}. The concept of the near-band-gap photoemission spectroscopy relies on the fact that the distribution of hot photoelectrons, which retain a part of their excess energy above thermal energy $kT$ prior to the emission into vacuum, contains a set of features related to the band structure of the photocathode active layer, as illustrated in Fig.~\ref{Fig.1}(a). Indeed, a small fraction of photoelectrons ballistically escape into vacuum without scattering in the active layer. Within the parabolic approximation the initial kinetic energy of photoexcited electrons $E_{0}$ in the CB of Na$_{2}$KSb equals:
\begin{equation} \label{Eq.1}
 E_{0}(\hbar\omega) = (\hbar\omega - E_{\text{g}})/(1 + m_{\text{e}} / m_{\text{h}}),
\end{equation}
where $m_{\text{e}}$ and $m_{\text{h}}$ are the effective masses of electrons and holes, respectively; $\hbar\omega$ is the photon energy. Additionally, hot photoelectrons tend to accumulate in the side valley minima of the CB during thermalization. Therefore, the evolution of the energy distributions of emitted electrons upon varying photon energy should reveal both the thresholds of interband transitions and the CB minima [see Fig.~\ref{Fig.1}(b)].

It should be noted that the accurate measurements of energy and angle distributions of photoelectrons emitted by NEA photocathodes in the low energy range ($< 1$\,eV) are tricky due to the high sensitivity of low-energy electron trajectories to electric fields produced by a relatively high contact potential difference between NEA photocathodes and other components of electron spectrometers \cite{Bradley1977, Lee2015, Ichihashi2018, Karkare2019}. The correct measurements of low-energy electron distributions usually require complex custom-made electron spectrometers \cite{Drouhin1985_1, Orlov2001, Karkare2019}. 

Compact energy electron spectrometers can be realized in vacuum tubes with planar geometry [see Fig.~\ref{Fig.2}(a)]. In such tubes the longitudinal component ($E_{\text{lon}}$) of the total electron kinetic energy ($E_{\text{k}}$) in vacuum is measured as
\begin{equation} \label{Eq.2}
 E_{\text{lon}} = E_{\text{k}}\text{cos}^{2}(\theta) = p^{2}_{\text{lon}}/(2m_{\text{0}}),
\end{equation}
where $\theta$ is the polar emission angle, $p_{\text{lon}}$ is the electron momentum component, normal to the photocathode surface, and $m_{\text{0}}$ is the free electron mass. The electron energy selection is made by applying a uniform retarding electric field between the photocathode and anode. These compact spectrometers proved to be effective in studying the ballistic photoemission from \textit{p}-GaAs(Cs,O) photocathodes \cite{Orlov2000, Bakin2003, Tereshchenko2017}, electron energy relaxation by the cascade emission of optical phonons in \textit{p}-GaN(Cs,O) \cite{Pakhnevich2004, Rozhkov2016}, as well as the energy band diagram and directional photoemission in Na$_{2}$KSb(Cs,Sb) photocathodes \cite{Rusetsky2021, Rusetsky2022, Rozhkov2024}. 

The advantages of energy spectroscopy in compact vacuum tubes include the simplicity of apparatus, high energy resolution ($\sim 1$\,meV) and high stability of photocathode surfaces (with lifetimes of about 10 years), though they are obtained at the cost of the additional broadening of the measured longitudinal energy distribution curves (EDCs) due to the angular spread of emitted electrons. Since our previous studies indicate that electrons in the multialkali Na$_{2}$KSb(Cs,Sb) photocathodes emit predominantly along the normal to the surface \cite{Rozhkov2024, Rozhkov2025}, the longitudinal energy distributions should closely resemble total kinetic energy distributions. 

The studies of the near-band-gap photoemission from \textit{p}-GaAs(Cs,O) NEA photocathodes proved that electron energy distributions contain a set of features related to the band structure of GaAs, as well as to the mechanisms of electron energy relaxation and emission into vacuum \cite{Drouhin1985_1, Terekhov1994, Orlov2000}. These opportunities have not been explored for studying Na$_{2}$KSb so far.

It should be noted that the distributions of low energy electrons, emitted from Na$_{2}$KSb-based photocathodes, were measured earlier at room temperature in Refs.~\cite{Ghosh1980, Beguchev1988}. However, the fine structure of electron distributions, associated with the photoemission of ballistic electrons, was not observed, and the Na$_{2}$KSb band structure parameters were not obtained, apparently, due to the thermal broadening of photoelectron energy distributions \cite{Drouhin1985_1, Orlov2000, Rozhkov2024}. 

In this work, the electronic band structure of Na$_{2}$KSb is studied by a combination of the near-band-gap photoemission spectroscopy of high-efficiency multialkali Na$_{2}$KSb(Cs,Sb) NEA photocathodes performed at $T = 80$\,K, along with the density functional theory (DFT) calculations of the Na$_{2}$KSb electronic band structure. The bulk band gap, spin orbit splitting, side valley positions of the conduction band were obtained from the evolution of energy distributions with varying photon energy. The DFT-calculated energies of the Na$_{2}$KSb band structure, along with the energies of the final electronic states of optical transitions, agree well with the experimental values. We also discuss the comparison of the hot electron dynamics and electronic band structures of Na$_{2}$KSb and GaAs.

\section{\label{Sec.2}Experimental and computational details}

Multialkali Na$_{2}$KSb(Cs,Sb) photocathodes were grown on glass substrates \cite{Rusetsky2021, Rozhkov2024}. The Na$_{2}$KSb film thickness was in the range of 80--140\,nm. The Na$_{2}$KSb surface was activated to the NEA state by the co-absorption of Cs and Sb. The luminous sensitivities of the produced Na$_{2}$KSb(Cs,Sb) films reached up to 850\,$\mu$A/lm at $T = 295$\,K. The details of the preparation of semitransparent multialkali Na$_{2}$KSb(Cs,Sb) photocathodes are presented in Refs.~\cite{Dolizy1982, Erjavec1997}. The compact vacuum photodiodes were fabricated by mounting the planar photocathodes and anodes parallel to each other at a distance of about 0.2\,mm on the opposite sides of a cylindrical alumina ceramics body [see Fig.~\ref{Fig.2}(a)]. The working diameter of the photodiodes was equal to 20\,mm. The semitransparent anodes were made by depositing 9\,nm of Pd on flat glass substrates \cite{Tereshchenko2025}. The measurements of the transmission, reflection and photoemission QE spectra, as well as the EDCs were performed on Na$_{2}$KSb(Cs,Sb) photocathodes in both reflection (R-mode) and transmission (T-mode) illumination geometries. The spectra, measured in the R-mode, were corrected by the anode transmission spectra (with an average transmission of approximately 0.4). The electron longitudinal energy distributions $N_{\text{e}}(E_{\text{lon}})$ were obtained by measuring the derivatives of the photocurrent-voltage characteristics $J_{\text{ph}}$ \cite{Orlov2000, Rozhkov2024}:
\begin{equation} \label{Eq.3}
N_{\text{e}}(E_{\text{lon}}) \sim dJ_{\text{ph}}(U)/dU.
\end{equation}
The width of an instrumental contour in the EDC measurements was kept at 20\% of the energy width of fine features in the EDCs. The measurements were performed in the temperature range of 80--295\,K.

The DFT calculations were carried out with the Vienna \textit{ab initio} simulation package (VASP) \cite{Kresse1993, Kresse1996}. For the exchange-correlation potential, the generalized gradient approximation (GGA-PBE) \cite{Perdew1996} was used to determine the equilibrium structure. The interaction between ion cores and valence electrons was described by the projector augmented-wave method \cite{Blochl1994, Kresse1999}, while the alkali metal $p$ orbitals were treated as valence electrons. The Slater-type DFT-1/2 self-energy correction method \cite{Ferreira2008, Ferreira2011} with a partially ionized antimony potential was applied to obtain the realistic electronic band structure. The spin-orbit coupling (SOC) was taken into account both in the structural relaxation and in the calculation of the electronic band structure.

\section{\label{Sec.3}Results}

\subsection{\label{SubSec.1}Photoemission quantum efficiency spectra of Na$_{2}$KSb(Cs,Sb) photocathodes at 80\,K}

\begin{figure*}
\includegraphics[width=1.0\linewidth]{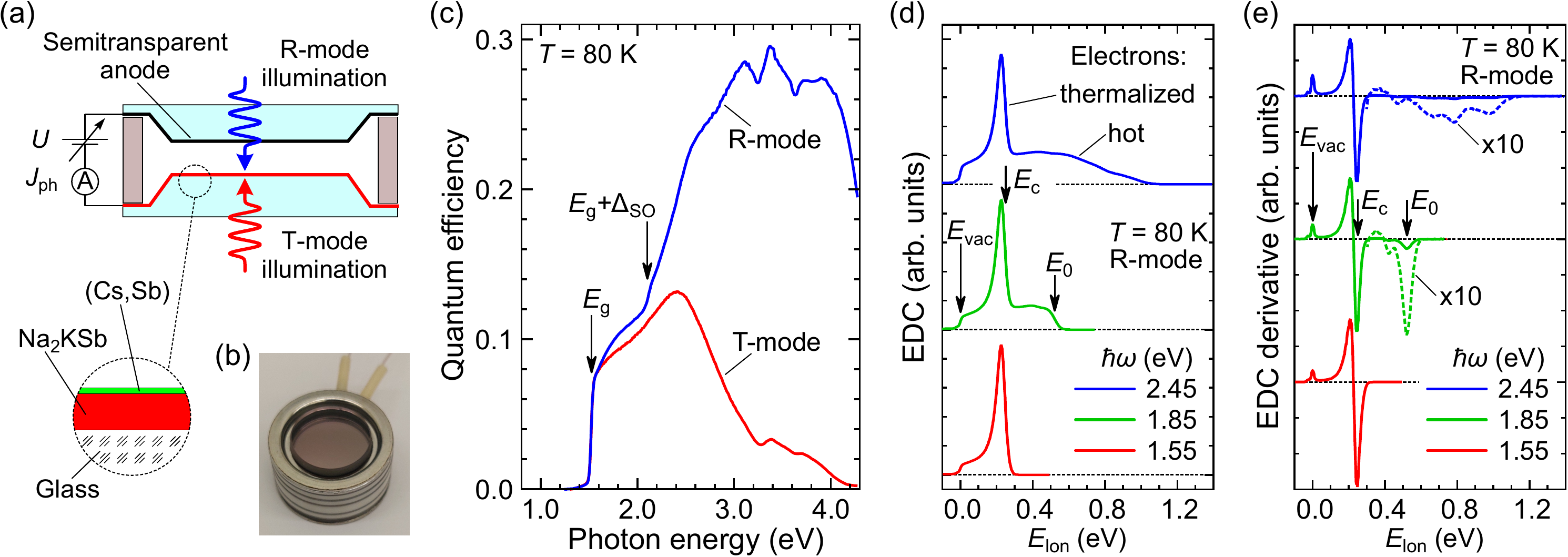}
\caption{\label{Fig.2} 
(a) Schematic cross-section and (b) photograph of a vacuum photodiode with the Na$_{2}$KSb(Cs,Sb) photocathode and a semitransparent anode. (c) Photoemission quantum efficiency spectra of the Na$_{2}$KSb(Cs,Sb) photocathode measured in transmission (T-mode) and reflection (R-mode) illumination geometries at $T = 80$\,K. The threshold energies for transitions from the valence bands to the conduction band ($E_{\text{g}}$ and $E_{\text{g}}+\Delta_{\text{SO}}$) are marked with arrows. The R-mode QE spectrum is normalized by the anode transmission spectrum. (d) Typical EDCs and (e) their derivatives of the Na$_{2}$KSb(Cs,Sb) photocathode measured in the R-mode at $T = 80$\,K. The hot and thermalized electron emission components are indicated for the EDC measured at $\hbar\omega = $2.45\,eV. The vacuum level $E_{\text{vac}}$, conduction band bottom in the bulk of Na$_{2}$KSb $E_{\text{c}}$ and the initial kinetic energy of the ballistic electrons in the conduction band $E_{\text{0}}$ are marked with arrows for EDC and its derivative measured at $\hbar\omega = 1.85$\,eV.
}
\end{figure*}

The QE spectra of a Na$_{2}$KSb(Cs,Sb) photocathode, measured in both T- and R-illumination geometries at $T = 80$\,K, are shown in Fig.~\ref{Fig.2}(c). At low $\hbar\omega$, the QE spectra are characterized by the sharp threshold of interband transitions in Na$_{2}$KSb with the band gap $E_{\text{g}}$ of about 1.52\,eV \cite{Rozhkov2024}. The positions of the threshold are determined by the maxima in the derivatives of the QE spectra. With increasing $\hbar\omega$, QE increases due to the increase in both the light absorption coefficient of the Na$_{2}$KSb layer and in the fraction of hot electrons, which emit with a higher escape probability \cite{Rozhkov2024}. The second threshold at $\hbar\omega = 2.11$\,eV is clearly seen in the R-mode QE spectrum. This threshold is due to the onset of optical transitions from the split-off (SO) valence band to the CB at the photon energy of $E_{\text{g}} + \Delta_{\text{SO}}$, where $\Delta_{\text{SO}}$ is the spin-orbit splitting. This yields $\Delta_{\text{SO}} = 0.59 \pm 0.04$\,eV. The non-monotonous variations of QE in the range of $\hbar\omega \approx$\,1.6--2.3\,eV are also caused by the optical interference in the Na$_{2}$KSb layer, which partially masks the threshold at $\hbar\omega = E_{\text{g}} + \Delta_{\text{SO}}$ in the T-mode QE spectrum, as is seen in Fig.~\ref{Fig.2}(c). It should be noted that both thresholds at $\hbar\omega = E_{\text{g}}$ and $\hbar\omega = E_{\text{g}} + \Delta_{\text{SO}}$ are clearly seen in the optical transmission spectra of the studied samples (not shown).

Further increase in the photon energy at $\hbar\omega > 2.40$\,eV leads to the decrease of the absorption length $\alpha^{-1}(\hbar\omega)$ of Na$_{2}$KSb layer to values, significantly smaller than the Na$_{2}$KSb layer thickness. In this case, photoelectrons are generated predominantly in the vicinity of the illuminated side of the Na$_{2}$KSb layer. For the R-mode illumination, electrons are generated near the emitting surface, which means that they remain in the Na$_{2}$KSb layer for only a short time before emission. This results in a reduced thermalization and recombination of electrons and, hence, in a large fraction of hot electrons in the emission current. This leads to the increase of QE with the increasing $\hbar\omega$ [see Fig.~\ref{Fig.2}]. On the contrary, in the case of T-mode illumination, electrons are generated near the glass-Na$_{2}$KSb interface, thus, requiring the maximum time to reach the emitting surface. This results in a stronger thermalization and recombination of electrons, and, thus, leads to a decrease in QE at a high $\hbar\omega$. The decrease of the QE above $\hbar\omega \approx 4.0$\,eV in both T- and R-mode QE spectra is caused by the optical absorption in the photocathode and anode glass substrates, respectively [see Fig.~\ref{Fig.2}(a)].

As is seen in Fig.~\ref{Fig.2}(c), the QE spectra contain various non-monotonous features in the $\hbar\omega$ range above approximately 2.4\,eV. The appearance of these features in both R- and T-mode QE spectra at close photon energies indicates that these features are related to the optical properties of the active Na$_{2}$KSb layer of the multialkali photocathode. The measurements of the transmission and reflection spectra showed that these features are caused by the variations of photocathode reflection coefficients, apparently, due to the variations of the optical joint density of states of Na$_{2}$KSb \cite{Rhim2005}. The analysis of these high energy features is beyond the scope of this work and will be presented elsewhere.

\subsection{\label{SubSec.2}Photoelectron energy distributions of Na$_{2}$KSb(Cs,Sb) photocathodes at 80\,K}

The longitudinal EDCs of the multialkali Na$_{2}$KSb(Cs,Sb) photocathode, measured in the R-mode illumination geometry at $T = 80$\,K, are shown in Fig.~\ref{Fig.2}(d). The EDCs are normalized to the maximum values. The EDCs, measured with photon energies exceeding the band gap ($\hbar\omega \geq E_{\text{g}}$), are formed by the emission of thermalized and hot photoelectrons \cite{Rozhkov2024}. At low photon energies ($\hbar\omega \approx E_{\text{g}}$) the EDCs are dominated by the emission of thermalized photoelectrons, which lose their excess energy above $kT$ before the emission into vacuum. The thermalized EDC component consists of a narrow peak near the CB bottom $E_{\text{c}}$ and a low-energy wing extending from $E_{\text{c}}$ down to the vacuum level $E_{\text{vac}}$, which is located about 250\,meV below $E_{\text{c}}$ \cite{Rozhkov2024}. At a larger $\hbar\omega$, the energy region extending from $E_{\text{c}}$ to the high energy edge of EDC $E_{0}$, is formed by the emission of hot photoelectrons \cite{Rozhkov2024}. It is seen in Fig.~\ref{Fig.2}(d) that, with the increasing $\hbar\omega$, the thermalized component of EDC remains almost unchanged, while the hot component extends toward higher energies, and its shape undergoes a complex transformation. It is convenient to study the weak fine structure of EDCs through the analysis of the EDC derivatives (DEDCs) \cite{Drouhin1985_1, Orlov2000, Rozhkov2024}, which are shown in Fig.~\ref{Fig.2}(e). Two major stationary (independent of $\hbar\omega$) peaks in DEDC correspond to the vacuum level $E_{\text{vac}}$ and the CB bottom in the bulk $E_{\text{c}}$ \cite{Rozhkov2024}. The hot component of DEDCs consists of multiple negative peaks with energies and amplitudes depending on $\hbar\omega$.

\begin{figure}
\includegraphics[width=0.78\linewidth]{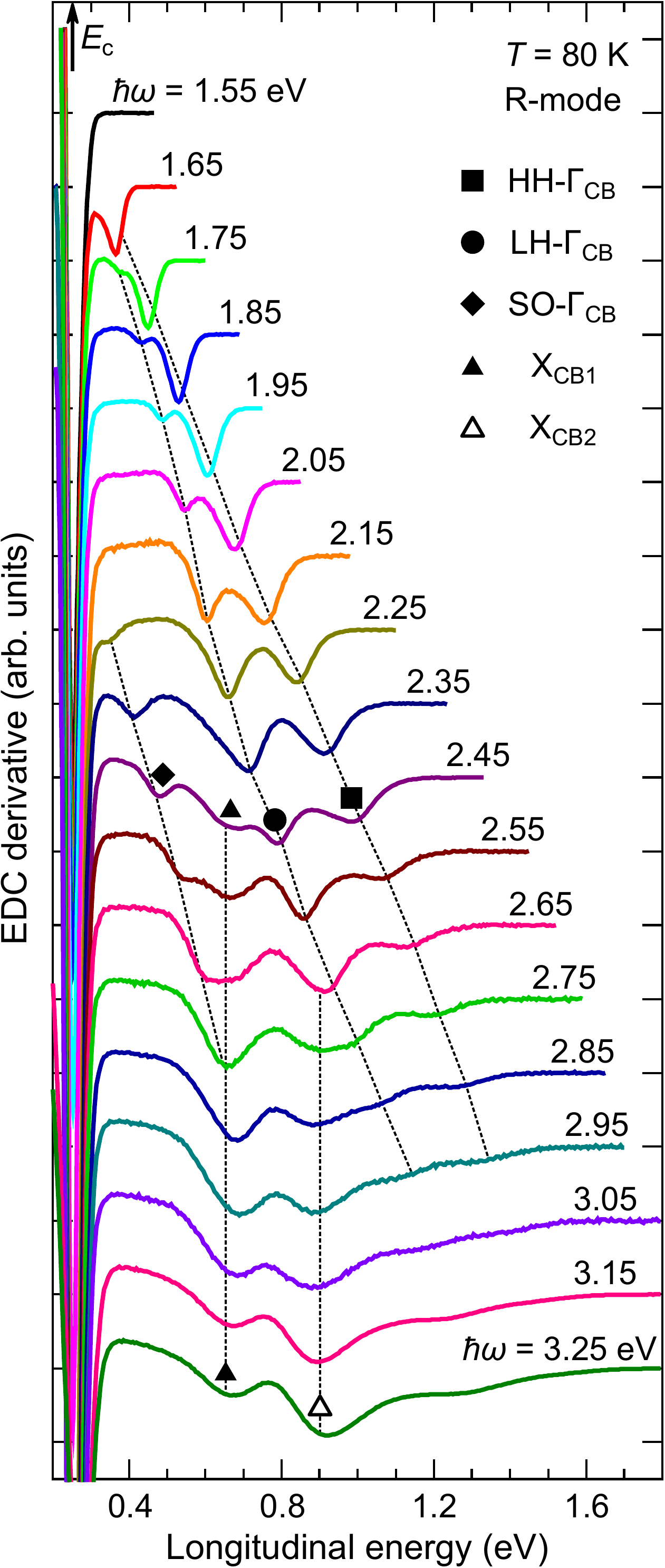}
\caption{\label{Fig.3}
The evolution of the EDC derivatives of the Na$_{2}$KSb(Cs,Sb) photocathode with varying $\hbar\omega$ measured in the R-mode at $T = 80$\,K. The derivatives are normalized to their maximum absolute values in the hot EDC region of $E_{\text{lon}} > 0.3$\,eV (the curves measured at $\hbar\omega = 1.55$\,eV and 1.65\,eV were normalized by the same value) and shifted vertically for clarity. Negative peaks corresponding to emission of ballistic electrons and hot electrons, scattered to the upper valleys of the conduction band, are marked for $\hbar\omega = 2.45$\,eV and $\hbar\omega = 3.25$\,eV: HH-$\Gamma_{\text{CB}}$(square), LH-$\Gamma_{\text{CB}}$(circle), SO-$\Gamma_{\text{CB}}$(diamond), $\text{X}_{\text{CB1}}$ (triangle) and $\text{X}_{\text{CB2}}$ (open triangle). The evolutions of peak positions with varying $\hbar\omega$ are highlighted with dashed lines as a guides to the eye. The position of the conduction band minima $E_{\text{c}}$ is marked with an arrow. 
}
\end{figure}

\begin{figure}
\includegraphics[width=1.0\linewidth]{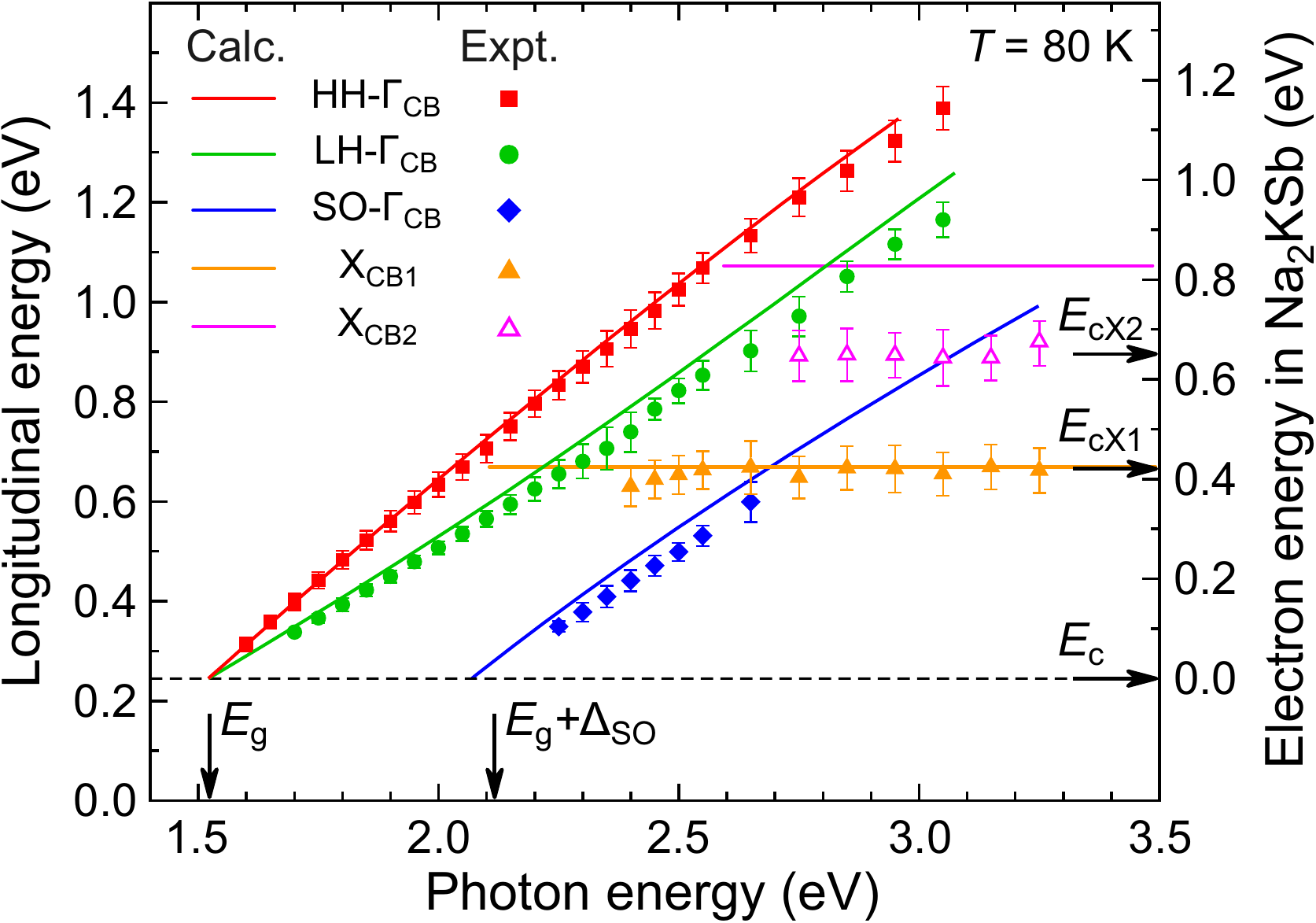}
\caption{\label{Fig.4} 
The evolution of the peak energy positions in EDC derivatives with varying $\hbar\omega$ for the Na$_{2}$KSb(Cs,Sb) photocathode measured at $T = 80$\,K. The DFT-calculated energies of ballistic electrons along the $\Gamma$-X direction and the bottoms of the upper valleys of the conduction band are shown with solid lines. The calculated curves are shifted toward the higher $\hbar\omega$ by 0.11\,eV to account for the difference between the calculated and measured band gaps (1.41\,eV and ($1.52 \pm 0.02$)\,eV, respectively). The experimentally determined threshold energies of optical transitions from valence bands to the conduction band ($E_{\text{g}}$ and $E_{\text{g}}+\Delta_{\text{SO}}$) and the positions of the conduction band minima ($E_{\text{c}}$, $E_{\text{c}\text{X1}}$ and $E_{\text{c}\text{X2}}$) are marked with arrows.
}
\end{figure}

The evolution of the hot component of EDC derivatives with varying $\hbar\omega$ is shown in Fig.~\ref{Fig.3}. The negative peaks of EDC derivatives can be divided into two groups. The peaks of the first group emerge from the dominating negative peak of the thermalized electrons located near the CB bottom $E_{\text{c}}$, when $\hbar\omega$ exceeds interband transition thresholds $E_{\text{g}}$ and ($E_{\text{g}} + \Delta_{\text{SO}}$). The energy positions of these peaks monotonously increase with the increasing $\hbar\omega$. We labeled these peaks as ``HH-$\Gamma_{\text{CB}}$", ``LH-$\Gamma_{\text{CB}}$" and ``SO-$\Gamma_{\text{CB}}$" and represented them with a filled square, circle and diamond, respectively, for the EDC derivative measured at $\hbar\omega = 2.45$\,eV. The peaks of the second group are stationary and appear at significantly higher photon energies. We labeled these stationary peaks as ``X$_{\text{CB}1}$" and ``X$_{\text{CB}2}$" and marked them with filled and open triangles, respectively, for the EDC derivatives measured at $\hbar\omega = 2.45$\,eV and 3.25\,eV. The evolution of the positions of those peaks with varying $\hbar\omega$ is highlighted in Fig.~\ref{Fig.3} by dashed lines. The peak positions in DEDC as a function of photon energy $\hbar\omega$ are shown in Fig.~\ref{Fig.4}.

First, consider the evolution of HH-$\Gamma_{\text{CB}}$, LH-$\Gamma_{\text{CB}}$ and SO-$\Gamma_{\text{CB}}$ peak energies with varying $\hbar\omega$. At $\hbar\omega$ slightly above $E_{\text{g}}$ (see the curve measured at $\hbar\omega = 1.55$\,eV in Fig.~\ref{Fig.3}), the hot electron component is masked by the strong negative peak of thermalized electrons. With the increasing $\hbar\omega$, the HH-$\Gamma_{\text{CB}}$ and LH-$\Gamma_{\text{CB}}$ peaks emerge and monotonously move toward higher electron energies. At photon energies $\hbar\omega < 2.0$\,eV, both peaks shift approximately proportional to $\hbar\omega$ with slope coefficients $A_{\text{HH}} = 0.83$ and $A_{\text{LH}} = 0.56$ for HH-$\Gamma_{\text{CB}}$ and LH-$\Gamma_{\text{CB}}$ peaks, respectively. The emergence of the two peaks just above the band gap of Na$_{2}$KSb and the linear dependences of the peak energies on $\hbar\omega$ with slopes lower than unity unambiguously indicate that HH-$\Gamma_{\text{CB}}$ and LH-$\Gamma_{\text{CB}}$ peaks correspond to the emission into vacuum of ballistic photoelectrons, which were excited to the CB from the heavy hole (HH) and light hole (LH) valence bands, respectively \cite{Drouhin1985_1, Orlov2000}. As can be seen in Fig.~\ref{Fig.4}, the extrapolated positions of the HH-$\Gamma_{\text{CB}}$ and LH-$\Gamma_{\text{CB}}$ peaks at low energies coincide with $E_{\text{c}}$ for $\hbar\omega$ = $E_{\text{g}}$, which indicates that the procedures of determining the band gap value $E_{\text{g}}$ and the position of $E_{\text{c}}$ are valid with an accuracy of about $\pm 20$\,meV \cite{Rozhkov2024}.

Upon a further increase in $\hbar\omega$ the SO-$\Gamma_{\text{CB}}$ peak emerges near $E_{\text{c}}$ (see the curve measured at $\hbar\omega = 2.25$\,eV in Fig.~\ref{Fig.3}). This peak also shifts linearly toward higher energies with the increasing $\hbar\omega$, with the slope coefficient $A_{\text{SO}} = 0.62$. Therefore, the SO-$\Gamma_{\text{CB}}$ peak is attributed to the emission of the ballistic photoelectrons which were excited to the CB from the SO valence band. As can be seen in Fig.~\ref{Fig.4}, the extrapolation of the SO-$\Gamma_{\text{CB}}$ peak energy to lower values shows that the SO-$\Gamma_{\text{CB}}$ peak is located at the bottom of the Na$_{2}$KSb CB when $\hbar\omega$ is equal to the second threshold in the QE spectra $E_{\text{g}} + \Delta_{\text{SO}} = 2.11$\,eV, which confirms the nature of this spectral threshold.

Now, consider the stationary X$_{\text{CB}1}$ and X$_{\text{CB}2}$ peaks. The X$_{\text{CB}1}$ peak becomes clearly visible when $\hbar\omega$ reaches approximately 2.4\,eV (see EDC derivative measured at $\hbar\omega = 2.45$\,eV in Fig.~\ref{Fig.3}), and its position $E_{\text{cX}1}$, which is equal to approximately $E_{\text{c}} + 0.41$\,eV, remains unchanged upon a further increase in $\hbar\omega$. A careful examination of the shapes of the EDCs derivatives indicates that the X$_{\text{CB}1}$ peak emerges actually at lower $\hbar\omega$. As can be seen in Fig.~\ref{Fig.4}, when $\hbar\omega$ reaches approximately 2.0\,eV and, consequently, when the energy of the HH-$\Gamma_{\text{CB}}$ peak reaches $E_{\text{cX}1}$, the ratio of HH-$\Gamma_{\text{CB}}$ peak and LH-$\Gamma_{\text{CB}}$ peak areas abruptly starts to decrease, and these peaks become asymmetric (see curves measured at $\hbar\omega = 2.05$\,eV and 2.15\,eV in Fig.~\ref{Fig.3}). Apparently, these changes are caused by the emergence of the X$_{\text{CB}1}$ peak at $E_{\text{cX}1} = E_{\text{c}} + 0.41$\,eV. 

The X$_{\text{CB}2}$ peak evolution is similar to that of the X$_{\text{CB}1}$ peak. The X$_{\text{CB}2}$ peak position $E_{\text{cX}2}$ is equal to approximately $E_{\text{c}} + 0.65$\,eV. While it has a well-defined peak shape only at $\hbar\omega$\,$> 2.80$\,eV (see curve measured at $\hbar\omega = 2.85$\,eV in Fig.~\ref{Fig.3}), again, the peak emergence occurs at a lower $\hbar\omega$\,$\approx 2.35$\,eV (see the curves measured at $\hbar\omega = 2.35$--2.55\,eV in Fig.~\ref{Fig.3}).

The emergence of X$_{\text{CB}1}$ and X$_{\text{CB}2}$ peaks, along with the changes in the HH-$\Gamma_{\text{CB}}$ and LH-$\Gamma_{\text{CB}}$ peak areas, can be readily explained by the onset of the strong intervalley scattering of the ballistic photoelectrons in the Na$_{2}$KSb CB, similarly to GaAs \cite{Drouhin1985_1, Fasol1990, Kanasaki2014}. Thus, we believe that the energies of the stationary X$_{\text{CB}1}$ and X$_{\text{CB}2}$ peaks correspond to the bottoms of two side valleys of the conduction band at the X point (see below subsection \ref{SubSec.3}); so, the side valley energy positions, with respect to the CB bottom in Na$_{2}$KSb $\Delta_{\Gamma-\text{X}1}$ and $\Delta_{\Gamma-\text{X}2}$, are equal to 0.41\,eV and 0.65\,eV, respectively.

At $\hbar\omega > 3$\,eV we also observed the emergence and evolution of other features in the hot component of the DEDCs. For example, a small negative peak at $E_{\text{lon}} \approx 1.3$\,eV appears on the curve measured at $\hbar\omega = 3.25$\,eV in Fig.~\ref{Fig.3}. Analysis of these high energy features is beyond the scope of the present work.

The evolution of the hot component of EDCs measured in the T-mode illumination geometry (not shown) is very close to that measured in the R-mode for $\hbar\omega < 2.4$\,eV. However, at a higher $\hbar\omega$, for the T-mode, the areas of the ballistic HH-$\Gamma_{\text{CB}}$, LH-$\Gamma_{\text{CB}}$ and SO-$\Gamma_{\text{CB}}$ peaks in EDC derivatives rapidly decrease with the increasing $\hbar\omega$ due to the increase in the photon absorption coefficient and, consequently, increase in the mean distance between the region of electron generation and the emitting surface. In particular, the stationary peak X$_{\text{CB}1}$ related to the first side valley of CB, is observed in both R- and T-modes. On the contrary, peak X$_{\text{CB}2}$ related to the second, higher lying side valley is not observed in the T-mode, presumably, due to a smaller light absorption length at larger photon energies $\hbar\omega>2.35$ eV.

Upon increasing the photocathode temperature, we observed a thermal broadening and blurring of the fine structure of DEDCs. This highlights the importance of studying the fine structure in DEDCs at cryogenic temperatures. The ballistic peak widths in the EDC derivatives also increase with the increasing $\hbar\omega$. At $T = 80$\,K the minimal widths FWHMs (full widths at half maximum) of the ballistic HH-$\Gamma_{\text{CB}}$, LH-$\Gamma_{\text{CB}}$ and SO-$\Gamma_{\text{CB}}$ peaks at low $\hbar\omega$ are equal to approximately 50\,meV and are determined by various physical factors, including possible chemical and doping inhomogeneities of the polycrystalline Na$_{2}$KSb layer along with the angular distribution of emitted hot electrons \cite{Rozhkov2024}. The quantitative analysis of the evolution of the peaks areas and widths with the increasing electron energy and $T$ is beyond the scope of this work and will be presented elsewhere.

\begin{table}
\caption{\label{Table.1} Measured and calculated energy gaps (in eV) in the Na$_{2}$KSb bulk band spectrum.}
\begin{ruledtabular}
\begin{tabular}{ccccc}
 & 
 $E_{\text{g}}$ & 
 $\Delta_{\text{SO}}$ &
 $\Delta_{\Gamma-\text{X}1}$ &
 $\Delta_{\Gamma-\text{X}2}$
 \\
\\ [-2.5ex]
\hline \\ [-2.0ex]
 & & Experiment & & \\
\\ [-2.5ex]
\hline \\ [-2.0ex]
This work & $1.52 \pm 0.02$ & $0.59 \pm 0.04$ & $0.41 \pm 0.05$ & $0.65 \pm 0.05$\\
\\ [-2.5ex]
\hline \\ [-2.0ex]
 & & Calculations & & \\
\\ [-2.5ex]
\hline \\ [-2.0ex]
This work & \multirow{2}*{1.41} & \multirow{2}*{0.55} & \multirow{2}*{0.42} & \multirow{2}*{0.82}\\
DFT-1/2 & &  &  & \\
GGA \cite{Yalameha2018} & 0.51 & 0.54 & 0.84 & 1.07\\
GGA \cite{Amador2021}  & 0.54 & 0.55 & 0.76 & 1.03\\
TB-mBJ \cite{Sharma2019} & 1.73 & 0.47 & 0.45 & 0.69\\
SCAN \cite{Amador2021}  & 1.11 & 0.54 & 0.63 & 0.92\\
$GW$ \cite{Amador2021}  & \multirow{2}*{1.51} & \multirow{2}*{--} & \multirow{2}*{0.77} & \multirow{2}*{1.04}\\
(wo-SOC)  &  &  &  & \\
\end{tabular}
\end{ruledtabular}
\end{table}

Thus, the measurements of the longitudinal EDCs, along with the optical and QE spectra, allowed us to reveal the details of the hot electron dynamics in Na$_{2}$KSb and to determine the band structure parameters of Na$_{2}$KSb. The experimentally obtained values of band gap $E_{\text{g}}$, spin-orbit splitting $\Delta_{\text{SO}}$ and the intervalley energy separations in the CB $\Delta_{\Gamma-\text{X}1}$ and $\Delta_{\Gamma-\text{X}2}$ are given in Table~\ref{Table.1}.

\subsection{\label{SubSec.3}The DFT-calculated electronic band structure of Na$_{2}$KSb}

\begin{figure*}
\includegraphics[width=1.0\linewidth]{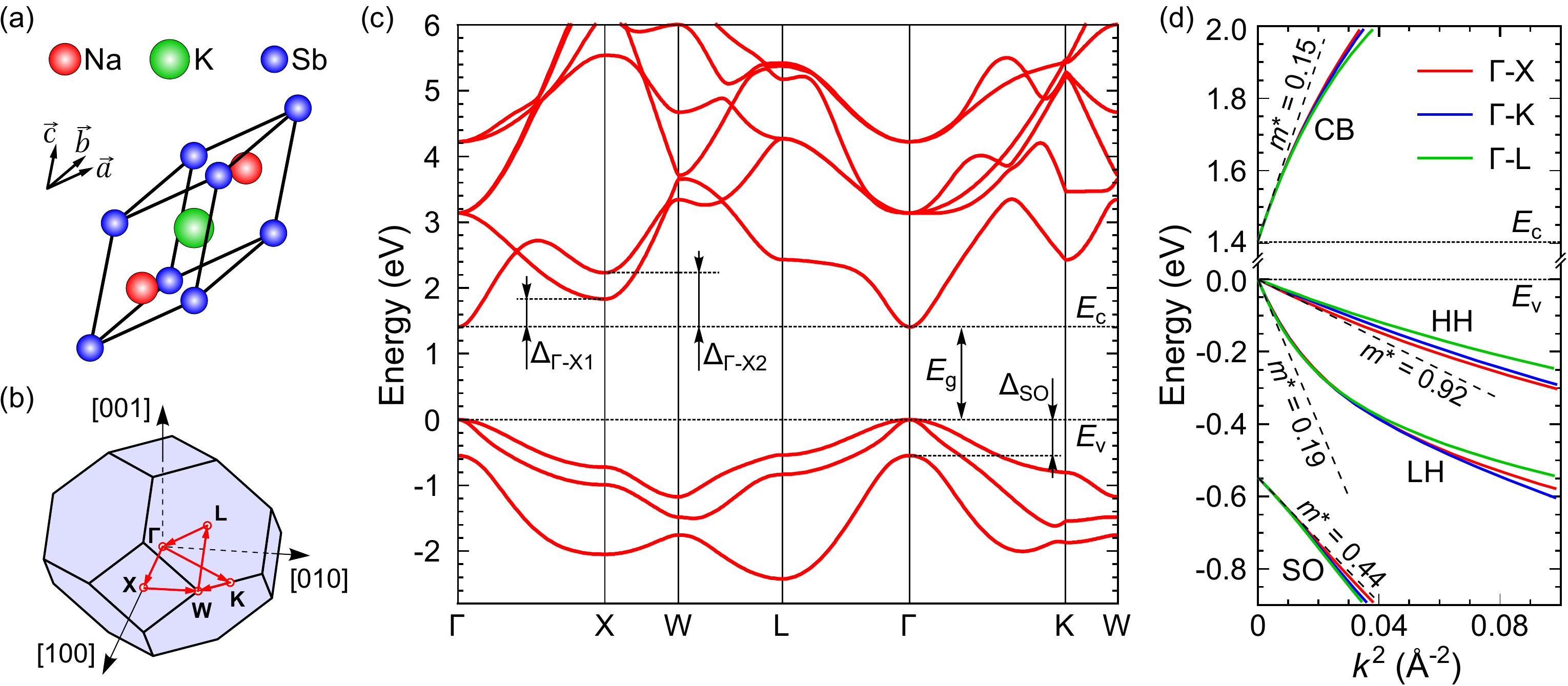}
\caption{\label{Fig.5} 
(a) Balls-and-sticks representation of the primitive cell of cubic Na$_{2}$KSb. (b) The Brillouin zone of Na$_{2}$KSb. The high-symmetry points and the path connecting them are highlighted. (c) The calculated band structure of Na$_{2}$KSb. $E_{\text{c}}$ and $E_{\text{v}}$ are conduction band bottom and valence band top, respectively. The energy gaps ($E_{\text{g}}$, $\Delta_{\text{SO}}$, $\Delta_{\Gamma-\text{X}1}$, $\Delta_{\Gamma-\text{X}2}$), are highlighted. (d) The calculated band structure of Na$_{2}$KSb at the center of the Brillouin zone along $\Gamma$-X, $\Gamma$-K and $\Gamma$-L directions, illustrating the bands non-parabalicity and anisotropy. The parabolic band approximations for the conduction band (CB), heavy hole (HH), light hole (LH) and split-off (SO) valence bands are shown by dashed lines for the $\Gamma$-X direction along with respective effective masses (in units of the free electron mass).
}
\end{figure*}

DFT calculations of the electronic structure of Na$_{2}$KSb, which took into account spin-orbit interaction, are presented in several papers \cite{Yalameha2018, Sharma2019, Amador2021}, although these papers are typically focused on the band structure near the Brillouin zone center. In our recent DFT calculation \cite{Rusetsky2022}, we also focused mainly on the fundamental band gap energy region in order to elucidate the symmetry of the wave functions in Na$_{2}$KSb and to explain the experimentally observed photoexcitation of spin-polarized electrons. To describe the experimental results on the band structure and dynamics of hot photoelectrons in Na$_{2}$KSb, which are presented in subsections \ref{SubSec.1} and \ref{SubSec.2}, here our calculations are aimed at a more comprehensive numerical evaluation of the Na$_{2}$KSb band structure, including the upper lying side valleys of the conduction band, and a realistic evaluation of the effective masses. Our calculations of the bulk spectrum carried out for a primitive unit cell [Fig.~\ref{Fig.5} (a)] along the high-symmetry directions of the corresponding Brillouin zone [Fig.~\ref{Fig.5} (b)] are presented in Fig.~\ref{Fig.5} (c). The notations of the quantities presented in Table~\ref{Table.1} are also provided there. Our calculated energies, alongside with those accurately extracted from the previously published spectra, are compared in Table~\ref{Table.1}. The calculated $E_{\text{g}}$ aligns well with the experimental value, whereas GGA calculations \cite{Yalameha2018,Amador2021} yield a band gap that is three times smaller than in the measurements. Meta-GGA calculations either overestimate (TB-mBJ \cite{Sharma2019}) or underestimate (SCAN \cite{Amador2021}) the band gap. The $GW$ method \cite{Amador2021} shows a perfect numerical agreement with the experimental results; however, these calculations do not account for SOC. At the same time, as can be seen in Table~\ref{Table.1}, the $\Delta_{\text{SO}}$ values are nearly independent of the calculation method, and our value, obtained in the quasiparticle approach, aligns well with those derived from earlier GGA and meta-GGA calculations, all matching experimental findings closely.

The positions of the stationary peaks X$_{\text{CB}1}$ and X$_{\text{CB}2}$ in the measured DEDCs, which we believe are responsible for the intervalley scattering of ballistic photoelectrons and are characterized by the energy distances between the bottom of the conduction band at $\Gamma$ and the side valleys, are in good agreement with the distances to the bottom of two valleys at the X point: $\Delta_{\Gamma-\text{X}1}$ and $\Delta_{\Gamma-\text{X}2}$ (Fig.~\ref{Fig.5} (c)). Our calculations for $\Delta_{\Gamma-\text{X}1}$ match the experimental value perfectly, while for $\Delta_{\Gamma-\text{X}2}$ they show a slight overestimation. Comparing the values of $\Delta_{\Gamma-\text{X}1}$ and $\Delta_{\Gamma-\text{X}2}$ extracted from earlier calculations with the experimental data, it is evident that only TB-mBJ demonstrates a satisfactory agreement, whereas other methods, including $GW$, yield significantly larger values, compared to the experiment. Therefore, we can conclude that the bulk spectrum obtained within DFT-1/2 provides the best representation of the experimentally derived band structure, and the assumption that intervalley transitions are determined by unoccupied bands at the X point is validated.

The electron and hole energies of Na$_{2}$KSb plotted \textit{versus} squared wavevector $k^{2}$ in the vicinity of the Brillouin zone center are shown in Fig.~\ref{Fig.5}(d) to illustrate the non-parabolicity and anisotropy of the electronic bands. Also, we show the parabolic approximations of the bands in the $\Gamma$-X direction by dashed lines with respective effective mass values in units of the free electron mass.

The calculated energies of the conduction band bottoms, as well as the calculated energies of the ballistic electrons, which are excited from the HH, LH and SO valence bands, are shown in Fig.~\ref{Fig.4} as solid lines. Only the energies of electrons exited along the $\Gamma$-X direction are shown since the valence band anisotropy is relatively weak in the energy range of interest.

\section{\label{Sub.4}Discussion}

The combination of the near-band-gap optical and photoemission spectroscopy of Na$_{2}$KSb(Cs,Sb) photocathodes, accompanied by the \textit{ab initio} calculations, revealed that both the hot electron dynamics and electronic band structure of Na$_{2}$KSb are qualitatively similar to those of GaAs \cite{Drouhin1985_1}. As we have already experimentally shown in our previous studies \cite{Rusetsky2021, Rusetsky2022, Rozhkov2024}, the band gap $E_{\text{g}}$ of Na$_{2}$KSb lies in the range of 1.50--1.55\,eV at $T = 80$\,K, which is close to the band gap of GaAs ($E_{\text{g}}^{\text{GaAs}} = 1.51$\,eV).

The obtained value of the Na$_{2}$KSb spin-orbit splitting $\Delta_{\text{SO}}$ of 0.59\,eV is by a factor of 2 higher than that of GaAs. This difference is in agreement with a slower reduction of the degree of circular photoluminescence polarization with the increasing $\hbar\omega$ in Na$_{2}$KSb, compared to GaAs, as well as with the high spin-polarization of photoelectrons emitted by Na$_{2}$KSb(Cs,Sb) photocathode \cite{Rusetsky2022}. Another consequence of a higher $\Delta_{\text{SO}}$ value in Na$_{2}$KSb, compared to GaAs, is that the non-parabolicity of the light hole valence band reveals itself at higher photon energies. As can be seen in Fig.~\ref{Fig.5}(d), the light hole valence band becomes parallel to the HH valence band, when the hole energy reaches $\Delta_{\text{SO}}$/2, similar to III-V semiconductors \cite{Drouhin1985_1}.

The evolution of the measured EDCs indicates that the hot electron dynamics in the Na$_{2}$KSb CB for electron energies above approximately 0.4\,eV is governed by the photoelectron scattering to the side valleys of the Na$_{2}$KSb conduction band, similar to GaAs. The examination of the calculated Na$_{2}$KSb band structure [see Fig.~\ref{Fig.5}(c)] shows that the two lowest side valleys in the CB are located at the X-point of the Brillouin zone, while in GaAs the first and second side valleys of the CB are located at the L-point and X-point, respectively \cite{Drouhin1985_1}.

It is seen in Fig.~\ref{Fig.5}(c) that another side valley of the Na$_{2}$KSb conduction band is located at the K-point of the Brillouin zone. Also, at the L-point there are CB states with a near-horizontal dispersion. Both K-point and L-point features of CB may manifest themselves in the energy distributions of hot electrons. A small negative peak in the EDC derivatives, located at the longitudinal energy of 1.3\,eV (see the curve measured at $\hbar\omega = 3.25$\,eV in Fig.~\ref{Fig.3}), could be related to one of these features of the Na$_{2}$KSb band structure. The analysis of the high energy regions of the band structure is beyond the scope of this work and will be presented elsewhere.

It should be noted that the observation of the EDC features, related to the X-valleys of the conduction band does not necessarily imply that electrons emit into vacuum directly from these X-valleys. Indeed, the direct photoemission from an X-valley could be forbidden due to the electron momentum conservation if the normal to the emitting surface of a Na$_{2}$KSb crystallite differs from the $\langle 100 \rangle$ crystallographic direction \cite{Bell1973, Drouhin1985_1}. In this case, photoelectrons, scattered into an X-valley, may first reenter the $\Gamma$-valley before emission. The orientation of the Na$_{2}$KSb crystallites in the multialkali photocathodes studied in this work is not exactly known. Previous studies of the Na$_{2}$KSb-based photocathodes grown on glass substrates showed that Na$_{2}$KSb crystallites are preferentially oriented in the photocathode plane. The predominant crystallographic plane of the Na$_{2}$KSb crystallites, parallel to the glass substrate, was reported to be $(111)$ \cite{Dolizy1988} or $(210)$ \cite{Ninomiya1969}. The (111) orientation was recently observed also on Na$_2$KSb grown on a graphene-coated SiC(0001) substrate \cite{Solovova2026}.

As can be seen in Fig.~\ref{Fig.5} (c) and Table~\ref{Table.1}, the DFT-calculated energies of the ballistic electrons in Na$_{2}$KSb, as well as side valley energies, are in good agreement with the measured peak energies in the EDC derivatives. The increase in slope in the calculated LH-$\Gamma_{\text{CB}}$ peak energy dependence on $\hbar\omega$ (see Fig.~\ref{Fig.4}), which is located at $\hbar\omega$ of about 2.3\,eV, is caused by a strong non-parabolicity of the LH valence band at the hole energies of about $\Delta_{\text{SO}}$/2 [see Fig.~\ref{Fig.5}(d)]. This increase in slope is also observed in the experimental data. Small differences between the calculated and measured energies, which are seen for ballistic electrons excited from the LH and SO valence bands (see LH-$\Gamma_{\text{CB}}$ and SO-$\Gamma_{\text{CB}}$ peaks in Fig.~\ref{Fig.4}), may be caused by the valence band splitting at the $\Gamma$-point. This splitting can be induced by a mechanical strain in the Na$_{2}$KSb due to the difference in the thermal expansion coefficients of the Na$_{2}$KSb polycrystalline film and the glass substrate. Also, it should be noted that the active layer of Na$_{2}$KSb(Cs,Sb) photocathode is, in fact, a solid solution Na$_{2-x}$K$_{1+x}$Sb \cite{McCarroll1960}; so, the observed differences may be partly caused by small variations of $x$ in the range of 0--0.05.

The dependence of the ballistic electron energy $E_{0}$ on $\hbar\omega$ can be used to obtain the ratios of the effective masses of electrons and holes in Na$_{2}$KSb, according to the parabolic bands approximation of $E_{0}(\hbar\omega)$ [see Eq.(\ref{Eq.1}) and Figs.~\ref{Fig.1}(a) and ~\ref{Fig.1}(b)]. Using the measured linear slopes $A_{\text{HH}}$, $A_{\text{LH}}$ and $A_{\text{SO}}$ of the HH-$\Gamma_{\text{CB}}$, LH-$\Gamma_{\text{CB}}$ and SO-$\Gamma_{\text{CB}}$ peak energy dependences $E_{0}(\hbar\omega)$, respectively, which were obtained in the electron energy range of 0.1--0.3\,eV relative to $E_{\text{c}}$ (see Fig.~\ref{Fig.4}), we determined $m_{\text{e}}/m_{\text{HH}} = 0.20 \pm 0.05$, $m_{\text{e}}/m_{\text{LH}} = 0.78 \pm 0.11$ and $m_{\text{e}}/m_{\text{SO}} = 0.61 \pm 0.09$, where $m_{\text{e}}$, $m_{\text{HH}}$, $m_{\text{LH}}$ and $m_{\text{SO}}$ are the effective masses of electrons, heavy holes, light holes and split-off holes in Na$_{2}$KSb, respectively. The measured ratios $m_{\text{e}}/m_{\text{HH}}$ and $m_{\text{e}}/m_{\text{LH}}$ agree, within the experimental error, with the respective calculated effective mass ratios, while the measured ratio $m_{\text{e}}/m_{\text{SO}}$ is by a factor of 2 larger than the calculated one [see Fig.~\ref{Fig.5}(d)]. A possible reason for this discrepancy could be the nonlinearity of $E_{0}(\hbar\omega)$ due to the non-parabolicity of the electron and hole dispersion curves. However, the analysis of the shapes of the calculated $E_{0}(\hbar\omega)$ curves (see Fig.~\ref{Fig.4}) showed that the variations of linear slopes $A_{\text{HH}}$, $A_{\text{LH}}$ and $A_{\text{SO}}$ are less than 10\% in the energy range of interest; so, the exact reason for the discrepancy between the experimental and calculated $m_{\text{e}}/m_{\text{SO}}$ ratios remains unclear.

It is seen in Fig.~\ref{Fig.3} that, at low photon energies $\hbar\omega < 2.0$\,eV, the area of the ballistic HH-$\Gamma_{\text{CB}}$ peak in the DEDCs of the Na$_{2}$KSb(Cs,Sb) photocathode exceeds the area of the LH-$\Gamma_{\text{CB}}$ peak by an order of magnitude. On the contrary, in the DEDCs of \textit{p}-GaAs(Cs,O) photocathodes, the areas of these peaks differ only by a factor of 2 \cite{Drouhin1985_1}. A possible reason for this difference consists in a significantly higher spin-orbit splitting in Na$_{2}$KSb and, consequently, later onset of the LH valence band non-parabolicity.

One can assume that various ballistic and stationary peaks, observed in the EDC derivatives of the polycrystalline multialkali Na$_{2}$KSb(Cs,Sb) photocathodes, can be related to different chemical phases of the material. Indeed, Na$_{3}$Sb, K$_{3}$Sb, NaK$_{2}$Sb compounds can be formed during the growth of active Na$_{2}$KSb layer under nonoptimal conditions \cite{McCarroll1960, Dowman1975, Dolizy1982}. However, these parasitic compounds are $n$-type semiconductors, while Na$_{2}$KSb is a $p$-type semiconductor \cite{McCarroll1971}. Consequently, due to the alignment of the Fermi level, the energies of the ballistic electrons in the $n$-type parasitic compounds are by approximately 1\,eV lower than in $p$-type Na$_{2}$KSb, and, therefore, are not detected in our near-band-gap photoemission experiments. Also, presumably, K$_{2}$CsSb and Cs$_{x}$Sb compounds can be formed at the surfaces of Na$_{2}$KSb crystallites during the photocathode activation with Cs and Sb \cite{Dowman1975, Galan1981, Dolizy1982, Erjavec1997}. However, the thicknesses of these activation layers were estimated to be of about 1\,nm, which are apparently too small to produce any observable features in the EDCs and their derivatives for the multialkali photocathodes with active Na$_{2}$KSb layer thicknesses of about 100\,nm. Again, the photocathodes studied in this work have high luminous sensitivities and QE values near the photoemission threshold $E_{\text{g}}$, which can be reached only with a high chemical homogeneity of the photocathode active layer and with a large size (about 100\,nm) of Na$_{2}$KSb crystallites \cite{Spicer1958, Ghosh1978, Rozhkov2024}. Finally, the evolution of various peaks in the EDCs derivatives of the Na$_{2}$KSb(Cs,Sb) NEA photocathodes with varying $\hbar\omega$ is well explained by the photoelectron excitation and thermalization processes in the homogeneous active layer of the multialkali photocathode.

The elimination of the uncertainties related to the chemical composition and crystallographic orientation of Na$_{2}$KSb crystallites requires a successful growth of the monocrystalline or ``fiber textured" \cite{Parzyck2023} Na$_{2}$KSb films on suitable substrates, along with \textit{ab initio} calculations of the Na$_{2}$KSb electronic band structure evolution upon variations in the alkali atom concentrations in a solid solution Na$_{2-x}$K$_{1+x}$Sb and upon variations in the mechanical strain in Na$_{2}$KSb films.

\section{\label{Sec.5}Conclusions}

In summary, we performed the near-band-gap photoemission spectroscopy on the polycrystalline multialkali Na$_{2}$KSb(Cs,Sb) NEA photocathodes, along with the \textit{ab initio} calculations of the Na$_{2}$KSb band structure. Our measurements of the energy distribution curves of Na$_{2}$KSb(Cs,Sb) photocathodes at various photon energies demonstrated that both the hot electron dynamics and electronic band structure of Na$_{2}$KSb exhibit qualitative similarities to those in GaAs. The ratios of the effective masses of electrons and holes, the band gap, spin-orbit splitting and side valley positions of the conduction band of Na$_{2}$KSb are obtained from the analysis of energy distributions and photoemission quantum efficiency spectra. The obtained values are in good agreement with the DFT-calculated band structure of Na$_{2}$KSb. The obtained results will advance the understanding of photoelectron and spin dynamics in Na$_{2}$KSb and will be useful for the development of robust spin-polarized electron sources based on Na$_{2}$KSb photocathodes.

\begin{acknowledgments}

The authors acknowledge the support from the Russian Science Foundation (Grant No. 25-62-00004) in the experimental part, SRF SKIF Boreskov Institute of Catalysis (FWUR-2024-0040) and ISP SB RAS. S.V.E. acknowledges the support from the Government research assignment for ISPMS SB RAS, Project No. FWRW-2026-0008.

\end{acknowledgments}

\bibliography{HotMultiPRB}

\end{document}